\documentclass[pdftex,twocolumn,epjc3]{svjour3}          

\RequirePackage[T1]{fontenc}

\smartqed  

\usepackage{graphicx}
\usepackage{flushend}
\usepackage[numbers,sort&compress]{natbib}
\usepackage[colorlinks,citecolor=blue,urlcolor=blue,linkcolor=blue]{hyperref}
\usepackage{widetext}
\usepackage{color}
\usepackage{dcolumn}
\usepackage{bm}
\usepackage{slashed}
\usepackage{amsmath}
\usepackage{latexsym}
\usepackage{amssymb}
\usepackage{mathrsfs}
\usepackage{amsfonts}
\usepackage{url}
\usepackage{mathtools}
\allowdisplaybreaks

\DeclarePairedDelimiter{\floor}{\lfloor}{\rfloor}

\journalname{Eur. Phys. J. C}

\begin{document}

\title{Applying the Gibbons-Werner method to bound orbits of massive particles in stationary spacetimes}


\author{Yang Huang\thanksref{e1,addr1,addr2}
        \and
        Xiangyun Fu\thanksref{addr1,addr2}
        \and
        Zhenyan Lu\thanksref{addr1,addr2}
        \and
        Xin Qin\thanksref{addr1,addr2}
        \and
        Xuan Zhou\thanksref{addr1,addr2}
        \and
        Lei Hui\thanksref{addr3}
}

\thankstext{e1}{corresponding author, yanghuang@mail.bnu.edu.cn}
\institute{Institute of Physics, Hunan University of Science and Technology, Xiangtan 411201, China \label{addr1}
           \and
           Key Laboratory of Intelligent Sensors and Advanced Sensing Materials of Hunan Province, Hunan University of Science and Technology, Xiangtan 411201, China \label{addr2}
           \and
           Peking University International S$\&$T Innovation Center at Lin-gang Special Area, Pilot Free Trade Zone, Shanghai 201306, China \label{addr3}
}

\date{Received: date / Accepted: date}

\maketitle

\begin{abstract}
    The Gibbons-Werner (GW) method provides a geometric framework for calculating the deflection angle of particles in curved spacetimes, and numerous extensions based on the original version have been developed in recent years to expand its applicability. Most existing studies, however, are restricted to unbound orbits. The finite-distance deflection angle, which assumes both the source and observer to be located at finite distances, motivates us to investigate the bending of bound orbits. In this work, we broaden the GW method to bound orbits of massive particles in stationary axisymmetric (SAS) spacetimes, following our previous extension in static spherically symmetric (SSS) backgrounds [Huang $et\ al.$, \href{https://doi.org/10.1103/PhysRevD.107.104046}{Phys. Rev. D 107, 104046 (2023)}]. By employing our generalized GW method for SAS spacetimes [Huang $et\ al.$, \href{https://doi.org/10.1088/1475-7516/2024/01/013}{J. Cosmol. Astropart. Phys. 01(2024)013}], (a) We obtain a formula for the deflection angle of bound massive particles in SAS spacetimes by dividing bound orbits that azimuthally overlap with themselves into multiple non-overlapping segments. This division enables the application of the GW method—originally developed for unbound orbits—to each segment in a consistent manner. (b) We overcome the limitation associated with the loss of positive definiteness of the Jacobi-Maupertuis Randers-Finsler (JMRF) metric, which occurs because bound massive particles have energies below unity. To show the practical implementation of our approach, we carry out the calculation in Kerr spacetime and obtain the deflection angle between two arbitrary points along the bound orbit of massive particles.
\end{abstract}
\maketitle

\section{Introduction}\label{sec-introduction}
Einstein’s general relativity is often summarized by the statement: "Spacetime tells matter how to move; matter tells spacetime how to curve". A natural implication of the first half of this statement is that the motion of particles—whether massless or massive—encodes information about the background spacetimes. In relativistic astrophysics, extracting observable quantities from particle trajectories in curved spacetimes is therefore a topic of central importance.

Among such observable quantities, the deflection angle plays a crucial role in understanding particle trajectories. Assuming that the photons propagate along null geodesics in the gravitational field of the Sun, Einstein calculated the deflection of light in 1916 \cite{einstein1916die}. The result was subsequently confirmed in 1919 by Dyson, Eddington, and Davidson through their observations of distant starlight during a solar eclipse \cite{dyson1920ix}, marking the first experimental test of general relativity. Since then, numerous investigations involving the photon deflection have emerged, such as the time delay of gravitational lensing \cite{treu2016time} and the shadow of black holes \cite{moffat2020shadow}. For massive particles, the deflection angle of unbound orbits has also received significant attention \cite{accioly2002gravitational,accioly2003gravitational,wucknitz2004deflection,bhadra2007testing,tsupko2014unbound,liu2016gravitational,he2016gravitational,he2017analytical,das2017motion,jusufi2019distinguishing,crisnejo2019higher,crisnejo2019gravitational,yang2019post,xiankai2019gravitational,he2020gravitational,huang2020perturbative,li2020gravitational,li2020circular,li2020finite,li2021deflection,li2021kerrnewman,li2022deflection,huang2023finite}. Moreover, for bound orbits, the pericenter advance angle can be regarded as the deflection angle between two successive pericenters \cite{einstein1915sitzungsberichtea,will2014confrontation}.

GW method is a powerful tool for calculating and understanding the deflection angle of particles from the geometric perspective. Although this method is initially proposed for infinite-distance deflection angle (both the source and observer are assumed at infinity) of photons in SSS spacetimes \cite{gibbons2008applications}, it is developed by researchers and now applicable for calculating the finite-distance and infinite-distance deflection angle of massless and massive particles in SSS and SAS spacetimes with or without asymptotic flatness \cite{werner2012gravitational,ishihara2016gravitational,ono2017gravitomagnetic,crisnejo2018weak,jusufi2018gravitational,li2020thefinitedistance,huang2024generalized}.

The model of bound orbits of massive particles serves as a fundamental framework in astrophysics, with applications ranging from Earth satellites \cite{lucchesi2010accurate,everitt2011gravity} and planets in the Solar System \cite{iorio2019calculation} to S-stars orbiting Sgr A* \cite{abuter2020detection}. Among their most notable applications, the pericenter advance provides valuable information about both the underlying spacetime and the orbiting body \cite{einstein1915erklarung,thorne1985laws,kraniotis2007periapsis,borka2012constraints,bambhaniya2021precession,bambhaniya2021shadows,wang2024periapsis}. In addition, phenomena such as the zoom-whirl orbit and the innermost stable circular orbit have been extensively investigated \cite{glampedakis2002zoom,levin2009dynamics,dubeibe2016geodesic,rana2019astrophysically,liu2019periodic}. Furthermore, in extreme mass-ratio inspiral systems, gravitational wave emissions constitute a key observational signature for probing the central massive object \cite{ni2024space,glampedakis2002zoom,glampedakis2002approximating,drasco2006gravitational,barausse2007circular,levin2008periodic,hinderer2008two,levin2009dynamics,fujita2009efficient,fujita2009analytical,rana2019astrophysically}.

Given the GW method can compute the finite-distance deflection angle of unbound particles, where both the starting and ending points of the orbit are assumed to lie at finite distances, we propose extending this method to the deflection angle for bound particles, where all points along the orbit remain at finite distances. Our recent work has accomplished this extension for SSS spacetimes \cite{huang2023extending}. In this paper, we further extend the GW method to the bound orbits of massive particles in SAS spacetimes. Specifically, by employing the generalized GW method for SAS spacetimes, we successfully transplant the technique developed for SSS spacetimes in Ref.~\cite{huang2023extending} to this broader setting, and explicitly calculate the deflection angle between two arbitrary points along the bound orbit of massive particles in Kerr spacetime.

This paper is organized as follows. Sec.~\ref{sec-2} provides a brief review of the generalized GW method for SAS spacetimes. In Sec.~\ref{sec-3}, we extend the GW method to bound massive particles in SAS spacetimes and discuss the corresponding observational implications. Sec.~\ref{sec-4} demonstrates the computational technique with the Kerr spacetime as an example. Finally, Sec.~\ref{sec-5} presents the conclusion. Throughout this paper, the spacetime signature ($-, +, +, +$) and the geometric units $G=c=1$ are adopted.

\section{Generalized GW method for SAS spacetimes}
\label{sec-2}
\subsection{JMRF metric}
For a stationary spacetime whose metric reads
\begin{equation}
  \mathrm{d}s^2 = g_{00}(\boldsymbol{x})\mathrm{d}t^2 + 2 g_{0i}(\boldsymbol{x})\mathrm{d}t\mathrm{d}x^i + g_{ij}(\boldsymbol{x})\mathrm{d}x^i \mathrm{d}x^j,
  \label{StationMetric}
\end{equation}
the corresponding JMRF metric is constructed as \cite{chanda2019jacobi}
\begin{equation}
  \mathrm{d} \hat{s} = \sqrt{\alpha_{ij} \mathrm{d}x^i \mathrm{d}x^j} + \beta_i \mathrm{d} x^i
\end{equation}
where
\begin{align}
  \alpha_{ij}\mathrm{d}x^i\mathrm{d}x^j = & \frac{m^{2}\left(\mathcal{E}^{2} + g_{00}\right)}{-g_{00}}\left( g_{ij} -\frac{g_{0i} g_{0j}}{g_{00}}\right)\mathrm{d}x^i \mathrm{d}x^j, \label{stationaryalphaij}\\
  \beta_{i}\mathrm{d}x^i = & -m\mathcal{E}\frac{g_{0i}}{g_{00}}\mathrm{d} x^{i}, \label{onform}
  \end{align}
in which $m$ is the rest mass of the particle, $\mathcal{E}$ is the energy per rest mass. Denoting the space equipped with Eq.~\eqref{stationaryalphaij} as $M^{(\alpha 3)}$, one can demonstrate that a geodesic in the four-dimensional stationary spacetime (Eq.~\eqref{StationMetric}) can be put in one-to-one correspondence with a curve $\gamma$ in $M^{(\alpha 3)}$, and the deviation between $\gamma$ and the geodesic in $M^{(\alpha 3)}$ is described by the one-form in Eq.~\eqref{onform}.

Furthermore, for a SAS spacetime whose metric reads
\begin{equation}
  \begin{aligned}
    \mathrm{d} s^2 =& g_{tt}\left(r,\theta \right) \mathrm{d} t^2 + g_{rr}\left(r,\theta \right)\mathrm{d}r^2 + g_{\theta\theta}\left(r,\theta \right)\mathrm{d}\theta^2  \\
    & + g_{\phi\phi} \left(r,\theta \right) \mathrm{d}\phi^2 + 2 g_{t\phi}\left(r,\theta \right)\mathrm{d}t\mathrm{d}\phi,
    \label{SASmetric}
  \end{aligned}
\end{equation}
a geodesic confined to the equatorial plane ($\theta=\pi/2$ and $\mathrm{d}\theta=0$) can be put in be put in one-to-one correspondence with a curve $\gamma$ in a two-dimensional space $M^{(\alpha 2)}$, equipped with the line element
\begin{equation}
  \begin{aligned}
    \mathrm{d}l^2 & = \alpha_{rr}\left(r\right) \mathrm{d}r^2 + \alpha_{\phi\phi}\left(r\right) \mathrm{d}\phi^2 \\
    & = \frac{m^2 \left( \mathcal{E}^2+g_{tt}\right)}{-g_{tt}}g_{rr} \mathrm{d}r^2  +  \frac{m^2 \left( \mathcal{E}^2+ g_{tt} \right)}{-g_{tt}}g_{\phi\phi} \mathrm{d}\phi^2,
  \end{aligned}
        \label{metricMalpha2}
\end{equation}
which follows directly from Eq.~\eqref{stationaryalphaij}. The deviation between $\gamma$ and the geodesic in $M^{(\alpha 2)}$ is described by a one-form
        \begin{equation}
            \beta_\phi\left(r\right) = -m\mathcal{E} \frac{g_{t\phi}}{g_{tt}}.
            \label{betaMalpha2}
        \end{equation}
Thus we can investigate the trajectory of particles in the equatorial plane of SAS spacetimes with the help of $M^{(\alpha 2)}$.

\subsection{Generalized GW method}
\label{GGWSAS}
The generalized GW method offers a simpler computational procedure and, moreover, can be applied to asymptotically non-flat spacetimes \cite{huang2024generalized}. Here we provide a brief introduction to this method for unbound particles in SAS spacetimes.

\begin{figure}[!ht]
  \centering
      \includegraphics[width=0.9\columnwidth]{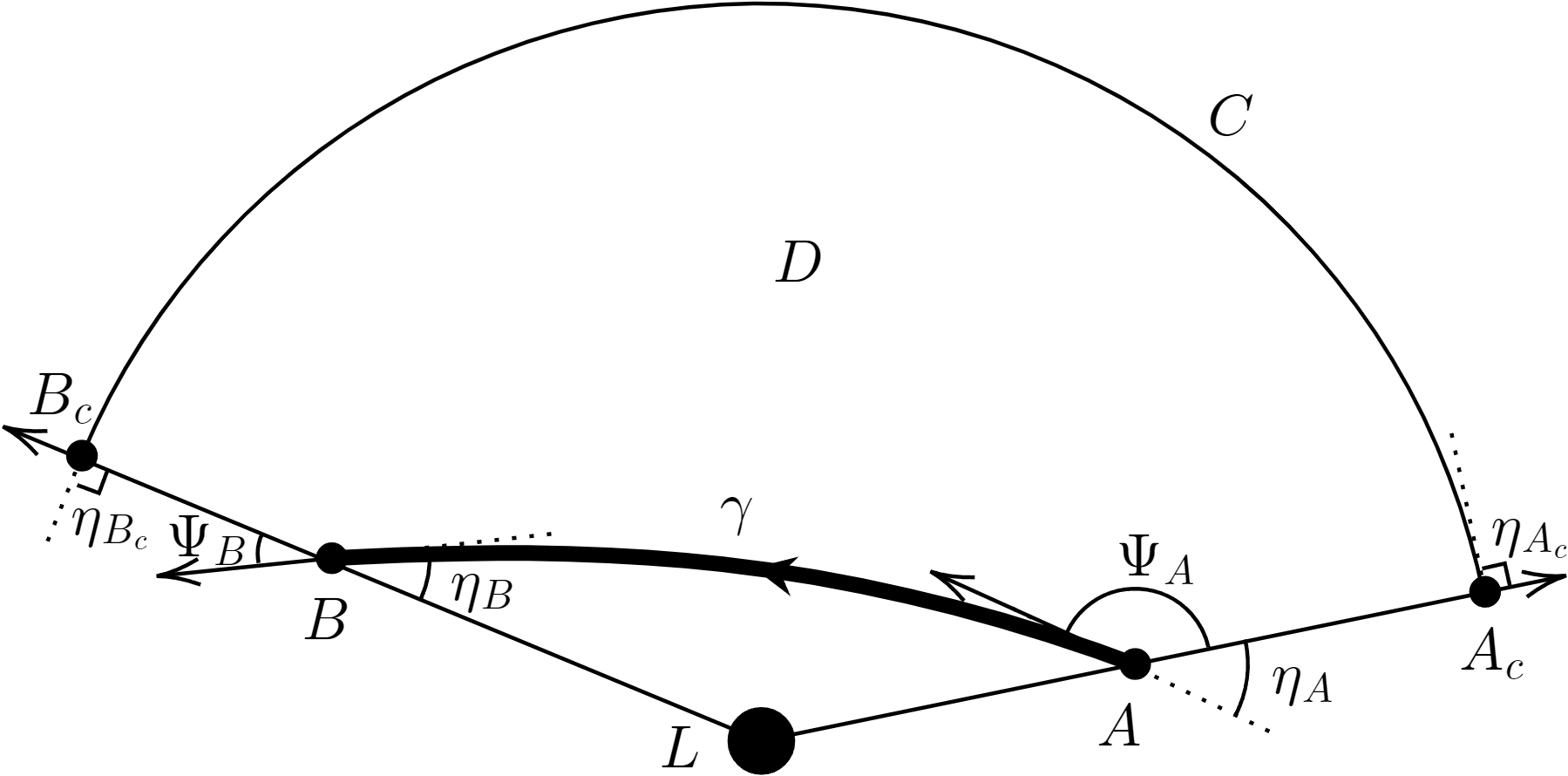} 
      \caption{The trajectory of unbound particles.}
  \label{fig-1}
\end{figure}

As shown in Fig.~\ref{fig-1},
in the $M^{(\alpha 2)}$ of a SAS spacetime, $\gamma$ is the trajectory of an unbound particle from the source (point $A$) to the observer (point $B$), $L$ is the lens, $\Psi_A$ and $\Psi_B$ are the angle between the tangent vector along $\gamma$ and the radial outward vector at $A$ and $B$, respectively. $C=\overset{\curvearrowright}{A_c B_c}$ is an arbitrary auxiliary circular arc whose radial coordinate is greater than the maximal radial coordinate of $\gamma$. $C$ intersects with the outgoing radial curves $\overrightarrow{LA}$ and $\overrightarrow{LB}$ at $A_c$ and $B_c$, respectively. $D=^{B_c}_B\square^{A_c}_A$ is a quadrilateral region surrounded by $\gamma$, $\overrightarrow{AA_c}$, $C$, and $\overrightarrow{BB_c}$. $\eta_A$, $\eta_{A_c}$, $\eta_{B_c}$ and $\eta_B$ are the exterior angle at vertices $A$, $A_c$, $B_c$, and $B$, respectively, measured in the counterclockwise orientation. Denoting the azimuthal coordinate of $A$ and $B$ as $\phi_A$ and $\phi_B$, respectively, the finite-distance deflection angle between $A$ and $B$ is defined as
\begin{equation}
 \delta_{BA}=\Psi_B-\Psi_A+\phi_B-\phi_A, 
  \label{DAdef}
\end{equation}
which was proposed and proved to be well-defined by Ishihara $et\ al$. \cite{ishihara2016gravitational}, and has since been widely employed \cite{ishihara2017finite,ono2017gravitomagnetic,ono2018deflection,ono2019deflection,haroon2019shadow,kumar2019shadow,crisnejo2019finite,ono2019effects,li2020finite,li2020thefinitedistance,takizawa2020gravitational,li2020circular,li2021kerr,li2021deflection,belhaj2022light,li2021kerrnewman,li2022deflection,pantig2022testing,huang2023finite}. Eq.~\eqref{DAdef} can also be rewritten as $\delta_{BA}=\Psi_{BA}+\phi_{BA}$, where $\Psi_{BA}=\Psi_B-\Psi_A$ represents the contribution from the velocity direction, and $\phi_{BA}=\phi_B-\phi_A$ corresponds to the contribution from the azimuthal position.

Applying the Gauss-Bonnet theorem (GBT, p.277 in Ref.~\cite{manfredo1976carmo}) to region $D$ yields
\begin{align}
  & \iint _{D} K\mathrm{d} S+\int _{\overrightarrow{AA_{c}}} \kappa \mathrm{d} l+\int _{C} \kappa \mathrm{d} l+\int _{\overrightarrow{B_{c} B}} \kappa \mathrm{d} l+\int _{\overset{\curvearrowright }{BA}} \kappa \mathrm{d} l \nonumber  \\
  & +\eta _{A} +\eta _{A_{c}} +\eta _{B_{c}} +\eta _{B} =2\pi \chi (D),
  \label{GBApply}
 \end{align}
in which $K$ represents the Gaussian curvature, $\kappa$ denotes the geodesic curvature, $\chi$ is the Euler characteristic number. Substituting $\kappa(\overrightarrow{AA_c})=\kappa(\overrightarrow{B_cB})=0$ (the proofs can be found in Appendix A of Ref.~\cite{huang2023finite}), $\int_{\overset{\curvearrowright}{BA}} \kappa \mathrm{d}l=-\int_\gamma \kappa \mathrm{d}l$, $\eta_A = \pi - \Psi_A$, $\eta_{A_c}=\eta_{B_c}=\pi/2$, $\eta_B=\Psi_B$, and $\chi\left(D\right)=1$ into Eq.~\eqref{GBApply} leads to
\begin{equation}
    \delta_{BA} = - \iint_{D} K \mathrm{d}S -\int_{C} \kappa \mathrm{d}l   +\phi_{B} - \phi_A+ \int_\gamma \kappa \mathrm{d}l.
    \label{deltageneral}
\end{equation}

In the generalized GW method \cite{huang2024generalized}, the integral of Gaussian curvature on $D$, i.e., the first term on the right-hand side of Eq.~\eqref{deltageneral}, is rewritten as
\begin{equation}
  \begin{aligned}
  \iint_{D} K \mathrm{d}S = & \int_{\phi_A}^{\phi_B} \int_{r_\gamma}^{r_c} K\sqrt{\alpha} \mathrm{d}r \mathrm{d}\phi \\
  = & \int_{\phi_A}^{\phi_B} \left[ H(r_c) - H(r_\gamma)  \right] \mathrm{d}\phi,
  \label{gaosiqulvDa}
\end{aligned}
\end{equation}
in which $r_c$ and $r_\gamma$ are respectively the radial coordinate of the auxiliary circular arc $C$ and the trajectory $\gamma$, and
\begin{equation}
  H(r)\equiv -\frac{\alpha_{\phi\phi,r}}{2\sqrt{\alpha}},  \ \  \int K\sqrt{\alpha} \mathrm{d}r  = H(r)+Cosnt.
  \label{Hr}
\end{equation}
In the derivation of Eq.~\eqref{gaosiqulvDa}, we adopt the metric of $M^{(\alpha 2)}$ in Eq.~\eqref{metricMalpha2}, together with the calculation formula of the Gaussian curvature \cite{werner2012gravitational}
\begin{equation}
  K=\frac{1}{\sqrt{\alpha}}\left[\frac{\partial}{\partial \phi}\left(\frac{\sqrt{\alpha}}{\alpha_{r r}} \Gamma_{r r}^\phi\right)-\frac{\partial}{\partial r}\left(\frac{\sqrt{\alpha}}{\alpha_{r r}} \Gamma_{r \phi}^\phi\right)\right],
  \label{GaussianCurvature}
\end{equation}
where $\alpha$ and $\Gamma$ are the determinant and Christoffel symbol of $M^{(\alpha 2)}$, respectively. Furthermore, the integral of geodesic curvature along $C$, i.e., the second term on the right-hand side of Eq.~\eqref{deltageneral}, can be reformulated as
\begin{equation}
  \int_{C}\kappa \mathrm{d}l = \int_{\phi_A}^{\phi_B} \left[ \kappa^{(c)} \frac{\mathrm{d}l}{\mathrm{d}\phi} \right]_{r=r_c} \mathrm{d}\phi = \int_{\phi_A}^{\phi_B} G(r_c) \mathrm{d}\phi,
   \label{kappajifen}
\end{equation}
in which $\kappa^{(c)}$ is the geodesic curvature along a circular arc, and
\begin{equation}
  G(r)  \equiv \kappa^{(c)} \frac{\mathrm{d}l}{\mathrm{d}\phi} = \frac{\alpha_{\phi\phi,r}}{2\sqrt{\alpha}}. 
  \label{Gr}
\end{equation}
Here we employ the metric of $M^{(\alpha 2)}$ in Eq.~\eqref{metricMalpha2} and the expression for the geodesic curvature of a circular arc in $M^{(\alpha 2)}$
\begin{equation}
  \kappa^{(c)} = -\Gamma^r_{\phi\phi}\frac{\alpha^{1/2}}{\alpha^{3/2}_{\phi\phi}}  = \frac{\alpha_{\phi\phi,r}}{2 \alpha_{\phi\phi}\sqrt{\alpha_{rr}}},
  \label{cediqulvcircular}
\end{equation}
which is derived from Liouville's formula for geodesic curvature (Chapter 4 of Ref.~\cite{struik1961lectures}).

Substituting Eqs.~\eqref{gaosiqulvDa} and \eqref{kappajifen} into Eq.~\eqref{deltageneral}, and noticing that $H(r_c)=-G(r_c)$ holds for any value of the radial of the auxiliary circular arc, we have
\begin{equation}
  \delta_{BA} =   \int_{\phi_A}^{\phi_B} \left[1+ H(r_\gamma)  \right] \mathrm{d}\phi + \int_\gamma \kappa \mathrm{d}l.
  \label{GGMdelta}
\end{equation}
Although Eq.~\eqref{GGMdelta} is derived based on the scenario $r_c > r_\gamma^{max}$, it also holds for the cases $r_\gamma^{max}\ge r_c\ge r_\gamma^{min}$ and $ r_c < r_\gamma^{min}$, where $r_\gamma^{max}$ and $r_\gamma^{min}$ are the maximum and minimum of the radial coordinates of $\gamma$, respectively \cite{huang2024generalized}. In addition, Eq.~\eqref{GGMdelta} can be further simplified into a more computationally convenient form
\begin{equation}
  \delta_{BA} = \int_{\phi_A}^{\phi_B} f(r_\gamma) \mathrm{d}\phi,
  \label{delta19}
\end{equation}
in which $f$ is function whose expression reads 
\begin{equation}
  f(r) 
  \equiv 1-\frac{\alpha_{\phi\phi,r}}{2\sqrt{\alpha}} -\beta_{\phi,r} \sqrt{ \frac{1}{\alpha_{\phi\phi}} \left( \frac{\mathrm{d}r}{\mathrm{d}\phi}\right)^2+ \frac{1}{\alpha_{rr}} } .
  \label{Tr}
\end{equation}

\section{Extending GW method to bound massive particles in SAS spacetimes}
\label{sec-3}
In this section, we first present the constraints imposed on the JMRF metric by bound massive particles; then, we extend the GW method, which is typically applied to unbound orbits, to bound orbits in SAS spacetimes; finally, we discuss the observational correspondence associated with the deflection angle for such orbits.

\subsection{Constraints on JMRF metric from bound orbits}
According to Eq.~\eqref{stationaryalphaij}, the JMRF metric is valid only when $\mathcal{E}^{2} + g_{00} > 0$. Considering the bound massive particles must satisfy $\mathcal{E}<1$, the radial coordinate of the JMRF metric for such particles in SAS spacetimes is confined in a finite region. Specifically, for the equatorial bound massive particles in Kerr spacetime, $g_{00}=-(1-2M/r)$ and thus the region $r\ge 2M/(1-\mathcal{E}^2)$ must be singular since $\mathcal{E}^{2} + g_{00} \le 0$. This restriction does not arise for unbound orbits.

Other GW methods require the radial coordinate of the auxiliary circular arc $r_c$ to be either infinite \cite{gibbons2008applications,werner2012gravitational,ishihara2016gravitational,ono2017gravitomagnetic,crisnejo2018weak,jusufi2018gravitational,li2020thefinitedistance} or fixed at a specific value \cite{takizawa2020gravitational,li2020circular}. These requirements pose significant challenges when extending such methods to the bound orbit of massive particles, as $r_c$ may fall within the singular region. In contrast, the generalized GW method remains valid for arbitrary values of $r_c$, allowing for a natural and straightforward extension to bound orbital configurations.

\subsection{Deflection angle of bound orbits}
The calculation formula in Eq.~\eqref{delta19} is derived based on unbound orbits, however, in this subsection, we will show that it is valid for bound orbits as well. Unlike the unbound orbit, a bound orbit overlaps with itself azimuthally. To reconcile this difference and cast the issue of bound orbits into a form analogous to that of unbound orbits, we divide the bound orbit into multiple segments, such that each segment does not overlap with itself azimuthally.

\begin{figure}[!ht]
  \centering
      \includegraphics[width=0.9\columnwidth]{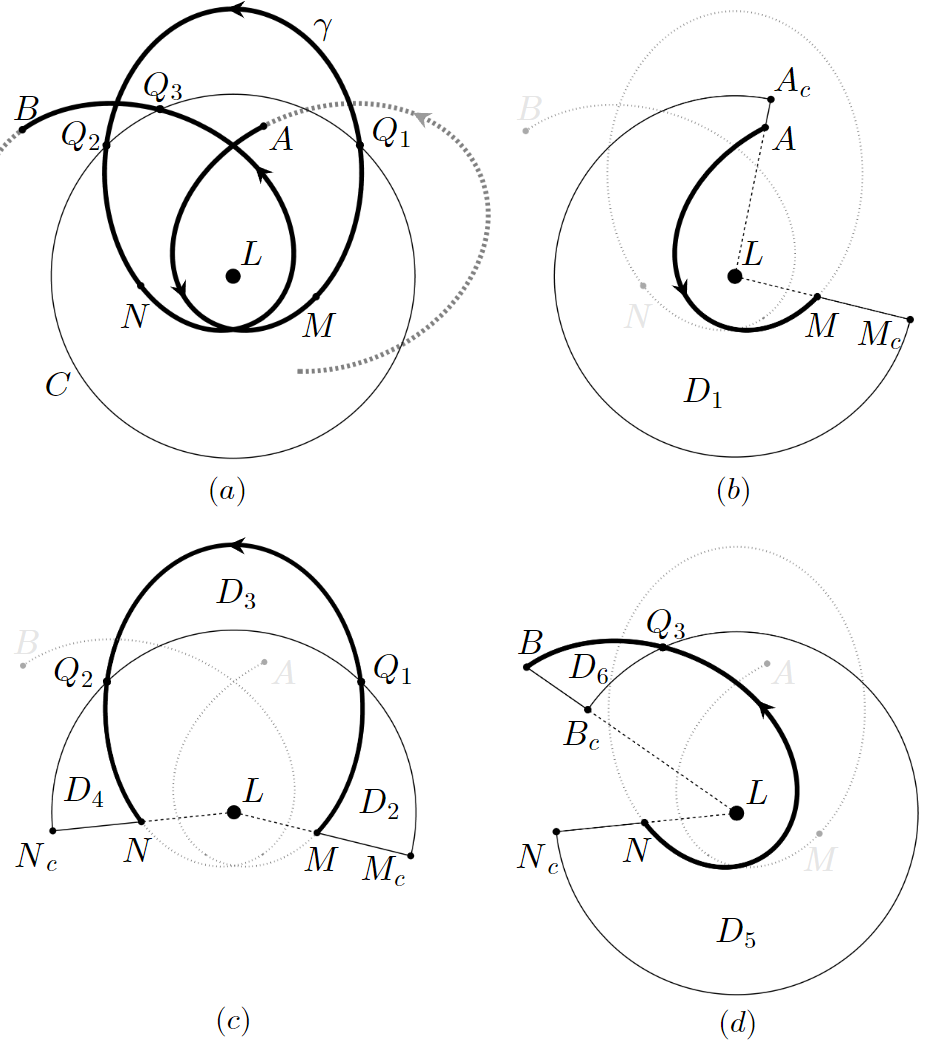} 
      \caption{The diagram for the trajectory of bound massive particles moving in the equatorial plane of SAS spacetimes.}
  \label{fig-2}
\end{figure}

As shown in Fig.~\ref{fig-2}, 
the subfigure (a) illustrates a part of the bound orbit of an equatorial massive particle in the SAS spacetime (assuming the particle move counterclockwise), $L$ denotes the lens, $C$ is the auxiliary circular arc, $A$ and $B$ are two arbitrary points along the orbit, $M$ and $N$ are two auxiliary points such that $\overset{\curvearrowright}{AM}$, $\overset{\curvearrowright}{MN}$, and $\overset{\curvearrowright}{NB}$ do not overlap with themselves azimuthally. We analyze the segments $\overset{\curvearrowright}{AM}$, $\overset{\curvearrowright}{MN}$, and $\overset{\curvearrowright}{NB}$ in subfigures (b), (c), and (d), respectively.

For the segment $\overset{\curvearrowright}{AM}$, the auxiliary circular arc $C$ intersects with the outgoing radial curves $\overrightarrow{LA}$ and $\overrightarrow{LM}$ at $A_c$ and $M_c$, respectively. Then we obtain a quadrilateral region $D_1=^{M_c}_{M}\square^{A_c}_A$. Similar to the scenario in Sec.~\ref{GGWSAS}, applying the GBT to $D_1$ brings about
\begin{equation}
  \delta_{MA} = \int_{\phi_A}^{\phi_M} f(r_\gamma) \mathrm{d}\phi.
  \label{deltama}
\end{equation}
For the segment $\overset{\curvearrowright}{MN}$, the auxiliary circular arc $C$ intersects with the outgoing radial curves $\overrightarrow{LM}$ and $\overrightarrow{LN}$ at $M_c$ and $N_c$, respectively, and intersects with the trajectory segment $\overset{\curvearrowright}{MN}$ at $Q_1$ and $Q_2$, respectively. Then we get two triangle regions $D_{2} = _{M}\triangle^{Q_1}_{M_c}$ and $D_{4} = _{N_c}\triangle^{Q_2}_{N}$, and a digon region $D_3$ with vertexes $Q_1$ and $Q_2$. Applying the GBT to $D_2$, $D_3$ and $D_4$ yields
\begin{equation}
  \begin{aligned}
  & \iint _{D_{2}} K\mathrm{d} S+\int _{\overset{\curvearrowright }{M_{c} Q_{1}}} \kappa \mathrm{d} l+\int _{\overset{\curvearrowright }{Q_{1} M}} \kappa \mathrm{d} l+\int _{\overrightarrow{MM_{c}}} \kappa \mathrm{d} l   \\
  & +\eta _{M} +\eta _{M_c} +\eta _{Q_1}  =2\pi \xi (D_2),   \label{d2}
  \end{aligned}
\end{equation}
\begin{equation}
  \begin{aligned}
  & \iint _{D_{3}} K\mathrm{d} S+\int _{\overset{\curvearrowright }{Q_{1} \gamma Q_{2}}} \kappa \mathrm{d} l +\int _{\overset{\curvearrowright }{Q_{2} CQ_{1}}} \kappa \mathrm{d} l  \\
  &  +\eta _{Q_1}+\eta _{Q_2}  =2\pi \chi (D_3),  \label{d3} 
  \end{aligned}
\end{equation}
  \begin{equation}
  \begin{aligned}
  & \iint _{D_{4}} K\mathrm{d} S+\int _{\overset{\curvearrowright }{Q_{2} N_{c}}} \kappa \mathrm{d} l+\int _{\overrightarrow{N_{c} N}} \kappa \mathrm{d} l+\int _{\overset{\curvearrowright }{NQ_{2}}} \kappa \mathrm{d} l  \\
  & +\eta _{Q_2} +\eta _{N_c} +\eta _{N}  =2\pi \chi (D_4). \label{d4}
  \end{aligned}
\end{equation}
In the above formulas, $\overset{\curvearrowright }{Q_{1} \gamma Q_{2}}$ is the curve from $Q_1$ to $Q_2$ along $\gamma$, $\overset{\curvearrowright }{Q_{2} C Q_{1}}$ is the curve from $Q_2$ to $Q_1$ along $C$; $\eta_M,\ \eta_{M_c},\ \eta_{Q_1},\ \eta_{Q_2},\ \eta_{N_c},\ \eta_N$ denote the exterior angles at vertices $M,\ M_c,\ Q_1,\ Q_2,\ N_c,\ N$, respectively, measured in the counterclockwise orientation. Let Eq.~\eqref{d2} $-$ Eq.~\eqref{d3} $+$ Eq.~\eqref{d4}, and using
\begin{equation}
\begin{aligned}
    & \iint _{D_2} K\mathrm{d} S - \iint _{D_3} K\mathrm{d} S + \iint _{D_4} K\mathrm{d} S  \\
  = & \int_{\phi_M}^{\phi_{Q_1}} \left[ H(r_c) - H(r_\gamma)  \right] \mathrm{d}\phi  \\
    & - \int_{\phi_{Q_1}}^{\phi_{Q_2}} \left[ H(r_\gamma) - H(r_c)  \right] \mathrm{d}\phi  \\
    & +\int_{\phi_{Q_2}}^{\phi_{N}} \left[  H(r_c) - H(r_\gamma)   \right] \mathrm{d}\phi  \\
  = & \int_{\phi_{M}}^{\phi_{N}} \left[  H(r_c) - H(r_\gamma)   \right] \mathrm{d}\phi,
\end{aligned}
\end{equation}
\begin{equation}
  \begin{aligned}
    & \int _{\overset{\curvearrowright }{M_{c} Q_{1}}} \kappa \mathrm{d} l-\int _{\overset{\curvearrowright }{Q_{2} CQ_{1}}} \kappa \mathrm{d} l+\int _{\overset{\curvearrowright }{Q_{2} N_{c}}} \kappa \mathrm{d} l\\
   = & \int _{\overset{\curvearrowright }{M_{c} Q_{1}}} \kappa \mathrm{d} l+\int _{\overset{\curvearrowright }{Q_{1} CQ_{2}}} \kappa \mathrm{d} l+\int _{\overset{\curvearrowright }{Q_{2} N_{c}}} \kappa \mathrm{d} l\\
   = & \int _{\overset{\curvearrowright }{M_{c} N_{c}}} \kappa \mathrm{d} l=\int _{\phi _{M}}^{\phi _{N}} G( r_{c})\mathrm{d} l
   \end{aligned}
\end{equation}
\begin{equation}
\begin{aligned}
  & \int _{\overset{\curvearrowright }{Q_{1} M}} \kappa \mathrm{d} l-\int _{\overset{\curvearrowright }{Q_{1} \gamma Q_{2}}} \kappa \mathrm{d} l+\int _{\overset{\curvearrowright }{NQ_{2}}} \kappa \mathrm{d} l\\
 = & \int _{\overset{\curvearrowright }{Q_{1} M}} \kappa \mathrm{d} l+\int _{\overset{\curvearrowright }{Q_{2} \gamma Q_{1}}} \kappa \mathrm{d} l+\int _{\overset{\curvearrowright }{NQ_{2}}} \kappa \mathrm{d} l\\
 = & \int _{\overset{\curvearrowright }{NM}} \kappa \mathrm{d} l=-\int _{\overset{\curvearrowright }{MN}} \kappa \mathrm{d} l,
 \end{aligned}
\end{equation}
$\eta_M=\pi-\Psi_M,\ \eta_N=\Psi_N,\ \eta_{M_c}=\eta_{N_c}=\frac{\pi}{2} $, and $\chi(D_2)=\chi(D_3)=\chi(D_4)=1$, we have
\begin{equation}
  \delta _{NM} =\int _{\phi _{M}}^{\phi _{N}}[ H(r_{\gamma } )+1]\mathrm{d} \phi +\int _{\overset{\curvearrowright }{M N}} \kappa \mathrm{d} l = \int _{\phi _{M}}^{\phi _{N}} f(r_{\gamma } )\mathrm{d} \phi .
  \label{deltanm}
\end{equation}
For the segment $\overset{\curvearrowright}{NB}$, by applying the GBT to two triangle regions $D_{5} = _{N}\triangle^{Q_3}_{N_c}$ and $D_{6} = _{B}\triangle^{Q_3}_{B_c}$, one can easily derive
\begin{equation}
  \delta _{BN} = \int _{\phi _{N}}^{\phi _{B}} f(r_{\gamma } )\mathrm{d} \phi.
  \label{deltabn}
\end{equation}

Finally, by using Eqs.~\eqref{deltama}, \eqref{deltanm}, and \eqref{deltabn}, we get the deflection angle for equatorial bound massive particles in SAS spacetimes
\begin{equation}
  \delta_{BA}=\delta_{MA}+\delta_{NM}+\delta_{BN} = \int _{\phi _{A}}^{\phi _{B}} f(r_{\gamma } )\mathrm{d} \phi,
  \label{eqDAbound}
\end{equation}
which is consistent in form with that for unbound orbits, i.e., Eq.~\eqref{delta19}.

\subsection{Correspondence to observations}
\begin{figure}[!ht]
  \centering
      \includegraphics[width=0.75\columnwidth]{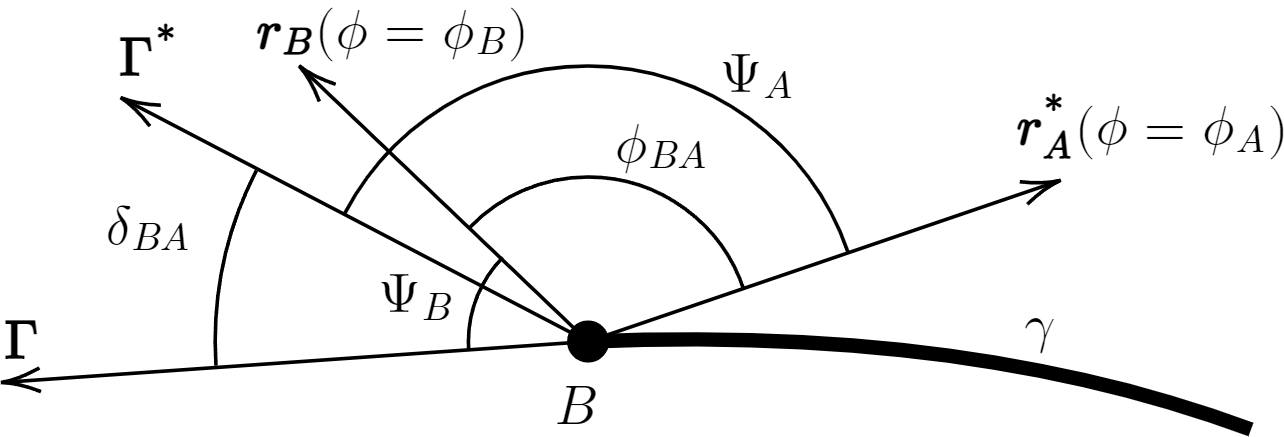} 
      \caption{The diagram for the interpretation of the finite-distance deflection angle at the observer.}
  \label{fig-3}
\end{figure}

We first review the interpretation to the finite-distance deflection angle defined by Eq.~\eqref{DAdef} from the perspective of the observer, as discussed by Takizawa, Ono, and Asada in Ref.~\cite{takizawa2020gravitational}. As shown in Fig.~\ref{fig-3},
focusing on the observer (point $B$) in Fig.~\ref{fig-1}, $\boldsymbol{\Gamma}$ represents the real light direction (the direction of light rays coming from the lensed image), $\boldsymbol{\Gamma^*}$ is the fiducial source direction (the direction of light rays coming from the unlensed source), the deviation between $\boldsymbol{\Gamma}$ and $\boldsymbol{\Gamma^*}$ is interpreted as the deflection angle defined by Eq.~\eqref{DAdef}. $\boldsymbol{r_B}$ denotes the radial direction at the observer, then the fiducial radial direction $\boldsymbol{r_A^*}$ can be determined by rotating the $\boldsymbol{r_B}$ with $\phi_{BA}=\phi_B-\phi_A$ which can be obtained from the ephemeris. By rotating the $\boldsymbol{r_A^*}$ with $\Psi_A$, which can be measured by observers at $B$ \cite{ono2019effects}, the $\boldsymbol{\Gamma^*}$ can be determined. Thus we have
\begin{equation}
  \Psi_B-\delta_{BA}=\Psi_A - \phi_{BA}
\end{equation}
at the position of the observer, i.e., $\delta_{BA}=\Psi_B-\Psi_A+\phi_{B}-\phi_A$. More details can be found in Ref.~\cite{takizawa2020gravitational}.

Following the above interpretation of the deflection angle in Eq.~\eqref{DAdef}, we extend this framework to bound orbits of massive particles, taking Mercury as an example. 

The orbital ephemeris of Mercury is determined not only by the Sun-Mercury gravitational interaction but also by perturbations from all other planets, the Earth-Moon system, and roughly 300 of the largest asteroids, as well as by the nonsphericity and tidal coupling between the Earth and the Moon. To extract the observational value of the pericenter advance for the inner planets Mercury, Venus, Earth, and Mars, Myles Standish, a senior researcher at the Jet Propulsion Laboratory (JPL), calculated their orbits over four centuries (from 1800 to 2200 A.D.) using the numerical integration program of the Solar System Data Processing System \cite{taylor2002exploring}. For each planet, two types of orbit were generated, one including relativistic corrections and the other based purely on Newtonian dynamics. By comparing the two, Standish obtained the pericenter advance attributable solely to general relativity. JPL maintains extensive databases to provide high-precision ephemeris including the positions and velocities for the planets and the Moon, with the Solar System Data Processing System \cite{jpl}. Comparable work has also been carried out by the Paris Observatory \cite{inpop} and the Institute of Applied Astronomy of the Russian Academy of Sciences \cite{epm}.

\begin{figure}[!ht]
  \centering
      \includegraphics[width=0.42\columnwidth]{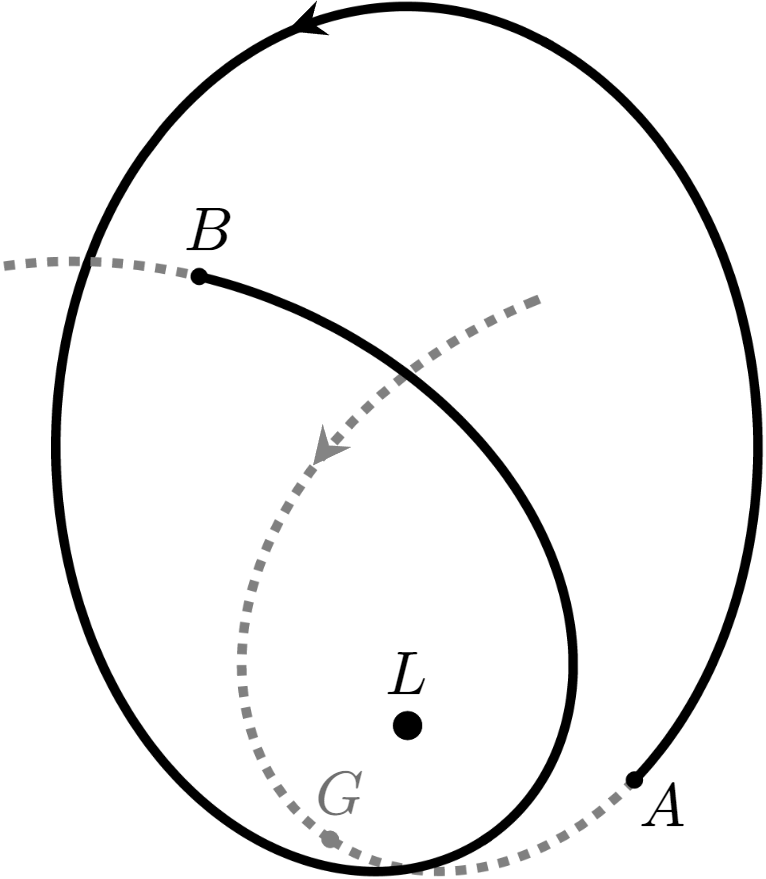} 
      \caption{The diagram for the trajectory of Mercury.}
  \label{fig-4}
\end{figure}

As illustrated in Fig.~\ref{fig-4},
we focus on the trajectory segment $\overset{\curvearrowright}{AB}$ of Mercury. By applying the framework of Fig.~\ref{fig-3} to point $B$ in Fig.~\ref{fig-4}, quantities such as $\Psi_B$ and $\phi_B$ at $B$ can be extracted from astronomical ephemeris, along with $\Psi_A$ and $\phi_A$ at point $A$. This allows the observed value $\delta_{BA}$ to be determined. In practice, $\delta_{BA}$ should be computed over many orbital periods so that statistical fluctuations are suppressed. By comparing relativistic and nonrelativistic results, the contribution to $\delta_{BA}$ from general relativity can be isolated.

The radial motion of Mercury (as well as other bound massive particles) is periodic and can be parameterized as \cite{chandrasekhar1983mathematical}
\begin{equation}
  r = \frac{p}{1+e\cos\xi},
  \label{rchi}
\end{equation}
where $e$ denotes the eccentricity and $p$ the semilatus rectum. The parameter $\xi$ increases monotonically with the orbital motion, thereby each value of $\xi$ uniquely identifies a point along the trajectory. Assigning $\xi=0$ to the pericenter $G$ in Fig.~\ref{fig-4}, then
\begin{equation}
  \begin{aligned}
  \delta_{BA}= & \Psi(\xi=2n\pi+\xi_B) - \Psi(\xi=2n\pi+\xi_A) \\
   &+ \phi(\xi=2n\pi+\xi_B)-\phi(\xi=2n\pi+\xi_A),
\end{aligned}
\end{equation}
which is averaged over multiple cycles ($n=0, \pm 1, \pm 2, \cdots$) for accuracy in actual calculation.

The pericenter advance angle $\Delta \omega$ depends only on azimuthal positions of two successive pericenters, while $\delta_{BA}$ involves both the azimuthal position and velocity direction of two arbitrary points. Hence $\delta_{BA}$ encodes more dynamical information than $\Delta \omega$, and the $\Delta \omega$ (plus $2\pi$) can be regarded as a special case of $\delta_{BA}$, i.e., $\Delta\omega + 2\pi = \delta_{BA}(\xi_A=0, \xi_B=2\pi)$, with $\Psi(\xi=2n\pi)=0$.

\section{Calculation in Kerr spacetime}
\label{sec-4}
In this section, we will show the calculation process of the deflection angle between two arbitrary points for bound massive particles in SAS spacetimes with the method proposed in this paper by taking the Kerr spacetime as an example. 

The Kerr metric with Boyer-Lindquist coordinates states \cite{boyer1967maximal}
\begin{equation}
    \begin{aligned}
        \mathrm{d} s^{2}=&-\left(1-\frac{2 M r}{\Sigma}\right) \mathrm{d} t^{2}-\frac{4 a M r \sin ^{2} \theta}{\Sigma} \mathrm{d} t \mathrm{d} \phi +\frac{\Sigma}{\Delta} \mathrm{d} r^{2}\\
        &+\Sigma \mathrm{d} \theta^{2}+\left[\Delta+\frac{2Mr\left(r^2+a^2\right)}{\Sigma}\right] \sin ^{2} \theta \mathrm{d} \phi^{2},
        \label{KerrMetric}
        \end{aligned}
\end{equation}
where $\Sigma = r^{2}+a^{2} \cos ^{2} \theta$, $\Delta=  r^{2}-2 M r+a^{2}$.

\subsection{Equation of motion}
For the massive particle in SAS spacetimes equipped with Eq.~\eqref{SASmetric}, two conserved quantities can be obtained from the Killing vectors $(\partial/\partial t)^a$ and $(\partial/\partial \phi)^a$ \footnote{The superscript “$a$” represents a vector index, which is distinct from the spin parameter $a$ appearing in Eq.~\eqref{KerrMetric}.}:
\begin{equation}
   \mathcal{E}  = -g_{tt}\frac{\mathrm{d}t}{\mathrm{d}\tau} -g_{t\phi} \frac{\mathrm{d}\phi}{\mathrm{d}\tau}, \ \ \ \  
   \mathcal{L}  = g_{t\phi} \frac{\mathrm{d}t}{\mathrm{d}\tau}+g_{\phi\phi} \frac{\mathrm{d}\phi}{\mathrm{d}\tau} , 
\label{EL}
\end{equation} 
where $\tau$ is the affine parameter of the timelike geodesic, $\mathcal{E}$ and $\mathcal{L}$ represent the energy and angular momentum per unit rest mass, respectively.

If the massive particle is further confined in the equatorial plane ($\theta=\pi/2, \mathrm{d}\theta=0$), the corresponding equation of motion can be derived by combing Eq.~\eqref{EL} and the normalization condition $g_{\mu\nu}(\mathrm{d}x^\mu/\mathrm{d}\tau)(\mathrm{d}x^\nu/\mathrm{d}\tau)=-1$:
\begin{align}
  & \frac{\mathrm{d} t}{\mathrm{d} \tau } =\frac{g_{t\phi }\mathcal{L} +g_{\phi \phi }\mathcal{E}}{g_{t\phi }^{2} -g_{tt} g_{\phi \phi }},\\
  & \frac{\mathrm{d} r}{\mathrm{d} \tau } =\pm \sqrt{\frac{g_{tt}\left( g_{\phi \phi } +\mathcal{L}^{2}\right) -g_{t\phi }^{2} +2g_{t\phi }\mathcal{EL} +g_{\phi \phi }\mathcal{E}^{2}}{g_{rr}\left( g_{t\phi }^{2} -g_{tt} g_{\phi \phi }\right)}}, \label{drdtauSASMP2}\\
  & \frac{\mathrm{d} \phi }{\mathrm{d} \tau } =\frac{g_{tt}\mathcal{L} +g_{t\phi }\mathcal{E}}{g_{tt} g_{\phi \phi } -g_{t\phi }^{2}}.
  \label{dphidtauSASMP2}
 \end{align}
According to Eqs.~\eqref{drdtauSASMP2} and \eqref{dphidtauSASMP2}, the trajectory equation of the equatorial massive particles can be obtained as
\begin{equation}
  \begin{aligned}
    \left(\frac{\mathrm{d} r}{\mathrm{d} \phi }\right)^{2} = & \frac{g_{t\phi }^{2} -g_{tt} g_{\phi \phi }}{g_{rr} (g_{tt}\mathcal{L} +g_{t\phi }\mathcal{E} )^{2}} \Big[ g_{tt}\left( g_{\phi \phi } +\mathcal{L}^{2}\right)\\
     & -g_{t\phi }^{2} +2g_{t\phi }\mathcal{EL} +g_{\phi \phi }\mathcal{E}^{2}\Big].
    \end{aligned}
    \label{drdphi2}
\end{equation}

\subsection{($\mathcal{E}$, $\mathcal{L}$) and ($e$, $p$)}
Substituting the Kerr metric Eq.~\eqref{KerrMetric} into Eq.~\eqref{drdtauSASMP2} leads to
\begin{equation}
  \begin{aligned}
    \left(\frac{\mathrm{d} r}{\mathrm{d} \tau }\right)^{2} = & \frac{1}{r^{3}}\Bigl[ (\mathcal{E}^{2} -1)r^{3} +2Mr^{2}\\
     & -\left( Y^{2} +a^{2} +2a\mathcal{E} Y\right) r+2MY^{2}\Bigr]
    \end{aligned}
  \label{drdtauKerr}
\end{equation}
where $Y=\mathcal{L}-a\mathcal{E}$. Let the right side of Eq.~\eqref{drdtauKerr} equal to zero, we get an equation
\begin{equation}
  (\mathcal{E}^2-1)r^3 + 2M r^2 - \left( Y^2 + a^2 + 2a\mathcal{E} Y \right) r + 2M Y^2 = 0,
  \label{requation}
\end{equation}
which shares the same roots with $\mathrm{d}r/\mathrm{d}\tau = 0$. According to Eq.~\eqref{rchi}, the radial coordinates of the pericenter and apocenter are
\begin{equation}
  r_1=\frac{p}{1+e}, \ \ \ \  r_2=\frac{p}{1-e},
\end{equation}
respectively. Both $r_1$ and $r_2$ correspond to the turning points where $\mathrm{d}r/\mathrm{d}\tau=0$ and, consequently, are the roots of Eq.~\eqref{requation}. Then Eq.~\eqref{requation} can also be written as
\begin{equation}
  (\mathcal{E}^2 - 1 )(r-r_1)(r-r_2)(r-r_3) = 0.
  \label{form2}
  \end{equation}
By comparing Eqs.~\eqref{requation} and \eqref{form2}, we have
\begin{align}
    & \mathcal{E} =  \sqrt{1+\frac{M(e^2-1)\left(p^2+e^2 Y^2-Y^2\right)}{p^3}}, \label{Eep}\\
    &   Y^2=\frac{-\mathcal{B}\pm \sqrt{\mathcal{B}^2 - 4\mathcal{A}\mathcal{C}}}{2\mathcal{A}} \label{Y2},
\end{align}
in which
\begin{equation}
  \begin{aligned}
    \mathcal{A} = & \left[ 1-\frac{2\left( 3+e^{2}\right) M}{p}\right]^{2} -\frac{4a^{2}\left( e^{2} -1\right)^{2} M}{p^{3}},\\
    \mathcal{B} = & 2\left( 3+e^{2}\right) M^{2} -2a^{2} -2Mp\\
     & -\frac{4a^{2}\left( e^{2} -1\right) M}{p} -\frac{2a^{2}\left( 3+e^{2}\right) M}{p},\\
    \mathcal{C} = & \left( a^{2} -Mp\right)^{2} .
    \end{aligned}
  \end{equation}
The "$+$" of the operator "$\pm$" in Eq.~\eqref{Y2} corresponds to the retrograde orbit, and "$-$" the prograde orbit \cite{bini2016gyroscope}. For the prograde orbit, we set $\mathcal{L}>0$ and $a>0$, thus
\begin{equation}
  \mathcal{L}_{pro}=Y_{pro}+a\mathcal{E}=\sqrt{\frac{-\mathcal{B}- \sqrt{\mathcal{B}^2 - 4\mathcal{A}\mathcal{C}}}{2\mathcal{A}} }+a\mathcal{E}.
    \label{Lp}
\end{equation}
For the retrograde orbit, we set $\mathcal{L}>0$ and $a<0$, thus
\begin{equation}
    \mathcal{L}_{retro}=Y_{retro}+a\mathcal{E}=\sqrt{\frac{-\mathcal{B}+ \sqrt{\mathcal{B}^2 - 4\mathcal{A}\mathcal{C}}}{2\mathcal{A}} }+a\mathcal{E}.
    \label{Lr}
\end{equation}

\subsection{Deflection angle}
Substituting the Kerr metric Eq.~\eqref{KerrMetric} into the metric of $M^{(\alpha 2)}$ in Eq.~\eqref{metricMalpha2} and the one-form in Eq.~\eqref{betaMalpha2} brings about
\begin{equation}
  \begin{aligned}
    \alpha _{rr} & =m^{2} r^{2}\left(\frac{\mathcal{E}^{2} r}{r-2M} -1\right)\frac{1}{r^{2} -2Mr+a^{2}}, \\
    \alpha _{\phi \phi } & =m^{2} r^{2}\left(\frac{\mathcal{E}^{2} r}{r-2M} -1\right)\frac{r^{2} -2Mr+a^{2}}{r^{2} -2Mr},\\
    \beta _{\phi } & =-\frac{2m\mathcal{E} Ma}{r -2M}.
    \end{aligned}
    \label{Kerralphabeta}
\end{equation}

Using the parameterized expression in Eq.~\eqref{rchi} for bound massive particles, we recast Eq.~\eqref{eqDAbound}, the calculation formula for the deflection angle of two arbitrary points for the equatorial massive particles in SAS spacetimes, as
\begin{equation}
  \delta_{BA} = \int_{\phi_A}^{\phi_B} f(r_\gamma) \mathrm{d}\phi =  \int_{\xi_A}^{\xi_B} z(\xi) \mathrm{d} \xi,
  \label{deltaBAbound}
\end{equation}
in which $z(\xi)=f(r_\gamma) \mathrm{d}\phi/\mathrm{d}\xi$. By substituting Eqs.~\eqref{drdphi2} and \eqref{Kerralphabeta} into Eq.~\eqref{Tr} and evaluating $\mathrm{d}\phi/\mathrm{d}\xi$ via Eqs.~\eqref{rchi} and \eqref{drdphi2}, together with the relations between ($\mathcal{E}$, $\mathcal{L}$) and ($e$, $p$) for prograde orbits given in Eqs.~\eqref{Eep} and Eq.~\eqref{Lp}, we obtain
\begin{equation}
  \begin{aligned}
    z(\xi )= & \frac{e\cos \xi +1}{e^{2} +2e\cos \xi +1} +M\cdot \frac{e\cos \xi +1}{p\left( e^{2} +2e\cos \xi +1\right)^{2}} \cdot \\
     & \left[ 4e\cos \xi \left( 2e^{2} +2e\cos \xi +3\right) +e^{4} +8e^{2} +3\right]\\
     & -M^{1/2} a\cdot \frac{4(e\cos \xi +1)}{p^{3/2}\left( e^{2} +2e\cos \xi +1\right)^{2}} \cdot \\
     & \left[ e^{2}\cos (2\xi )+2\left( e^{2} +2\right) e\cos \xi +4e^{2} +1\right]\\
     & +a^{2} \cdot \frac{e\cos \xi +1}{2p^{2}\left( e^{2} +2e\cos \xi +1\right)^{2}} \cdot \\
     & \left[ 2e^{2}\cos (2\xi )+4\left( e^{2} +3\right) e\cos \xi -e^{4} +12e^{2} +3\right]\\
     & +M^{2} \cdot \frac{e\cos \xi +1}{2p^{2}\left( e^{2} +2e\cos \xi +1\right)^{3}}\cdot\mathcal{I} +o\left( M^{2} ,a^{2}\right),
    \end{aligned}
    \label{zchi}
\end{equation}
where $\mathcal{I}$ is presented in the footnote \footnote{$\mathcal{I} =7e^{4}\cos (4\xi )+( 21e^{2} +59) e^{3}\cos (3\xi )+20e^{6} +258e^{4} +252e^{2} \\ +27+( 21e^{4} +176e^{2} +167) e^{2}\cos (2\xi )+( 3e^{6} +170e^{4} +510e^{2} +165) e\cos \xi $}. Finally, substituting Eq.~\eqref{zchi} into Eq.~\eqref{deltaBAbound} leads to the deflection angle between two arbitrary points for the bound massive particle in Kerr spacetime
\begin{equation}
  \delta_{BA}  = Z(\xi_B)-Z(\xi_A) + \floor*{\frac{\xi_B-\xi_A}{2\pi}} \pi,
  \label{deltaBAfinal}
\end{equation}
where $Z(\xi)$ denotes the antiderivative of $z(\xi)$ (the integration constant is omitted)
\begin{equation}
  \begin{aligned}
    Z(\xi )= & \frac{\xi }{2} +\arctan\left(\frac{e+1}{e-1}\cot\frac{\xi }{2}\right)\\
     & +M\cdot \frac{1}{p}\left[\frac{e\sin \xi \left( e^{2} +4e\cos \xi +3\right)}{e^{2} +2e\cos \xi +1} +3\xi \right]\\
     & -M^{1/2} a\cdot \frac{4}{p^{3/2}}\left[\frac{e\sin \xi (e\cos \xi +1)}{e^{2} +2e\cos \xi +1} +\xi \right]\\
     & +a^{2} \cdot \frac{1}{2p^{2}}\left[\frac{e\sin \xi \left( 2e\cos \xi +3-e^{2}\right)}{e^{2} +2e\cos \xi +1} +3\xi \right]\\
     & +M^{2} \cdot \frac{\mathcal{J}}{8p^{2}\left( e^{2} +2e\cos \xi +1\right)^{2}} +  o\left( M^{2} ,a^{2}\right),
    \end{aligned}
    \label{Zchi}
\end{equation}
and the quantity $\mathcal{J}$ is given in the footnote \footnote{$  \mathcal{J} =20( 5e^{4} +23e^{2} +6) e\sin \xi +24( e^{4} +19e^{2} +18) e\xi \cos \xi \\  +4e^{2}\cos (2\xi )( 3e^{2} \xi +7e^{3}\sin \xi +59e\sin \xi +7e^{2}\sin \xi \cos \xi +54\xi )\\   +6( e^{6} +22e^{4} +73e^{2} +18) \xi +( 3e^{4} +220e^{2} +235) e^{2}\sin (2\xi )$}. The symbol "$\floor{\ }$" in Eq.~\eqref{deltaBAfinal} denotes the floor function, which rounds a real number down to the nearest integer. The existence of the "$\floor{\ }$" term in Eq.~\eqref{deltaBAfinal} arises from the omission of the constant term in $Z(\xi)$ and the fact that the second term of $Z(\xi)$, namely, $Z^{(2)}\equiv\arctan[(e+1)\cot(\xi/2)/(e-1)]$, is a $2\pi$-period function with discontinuities at $\xi=2n\pi$ (specifically $\lim_{\xi \to 2n\pi^-} Z^{(2)}=\pi/2$ and $\lim_{\xi \to 2n\pi^+} Z^{(2)}=-\pi/2$).

It should be noted that although Eqs.~\eqref{deltaBAfinal} and \eqref{Zchi} are derived under the configuration of prograde orbits ($\mathcal{L}>0$ and $a>0$), our calculations indicate that the results of retrograde orbits ($\mathcal{L}>0$ and $a<0$) are fully consistent with them in form. In addition, if $A$ and $B$ are two successive pericenters, then $\delta_{BA}-2\pi$ gives the pericenter advance angle, which agrees with the results obtained by Weinberg \cite{weinberg1972principles} and Subramaniam \cite{SubramaniamKerr}, where the expressions consist of the constant term, $M$ term, and $M^{1/2}a$ term.

\begin{figure}[!ht]
  \centering
      \includegraphics[width=0.9\columnwidth]{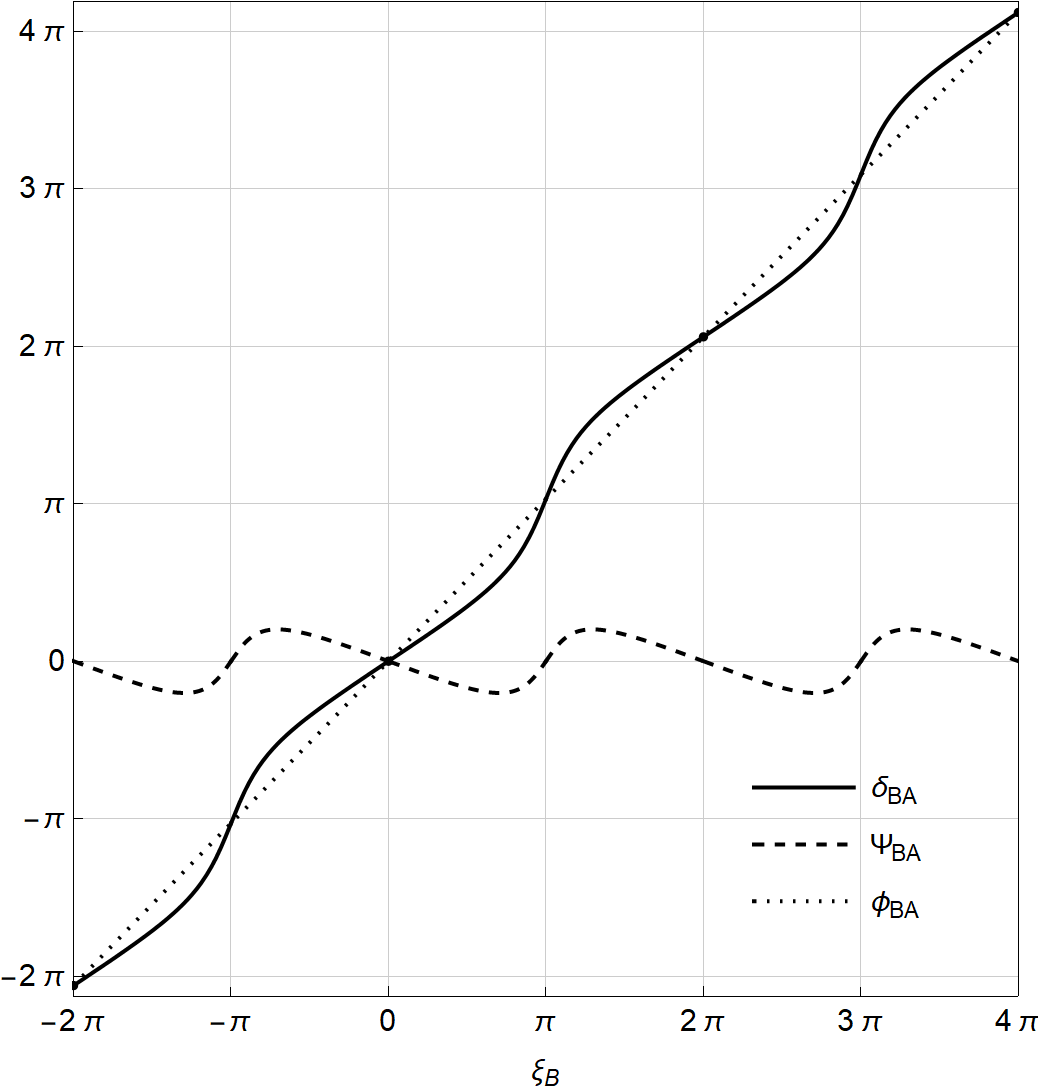} 
      \caption{$\delta_{BA}$, $\phi_{BA}$, and $\Psi_{BA}$ against $\xi_B$ with $\xi_A=0$, $a=0.4M$, $e=0.5$, and $p=10^2M$.}
  \label{fig-5}
\end{figure}

\begin{figure}[!ht]
  \centering
      \includegraphics[width=0.94\columnwidth]{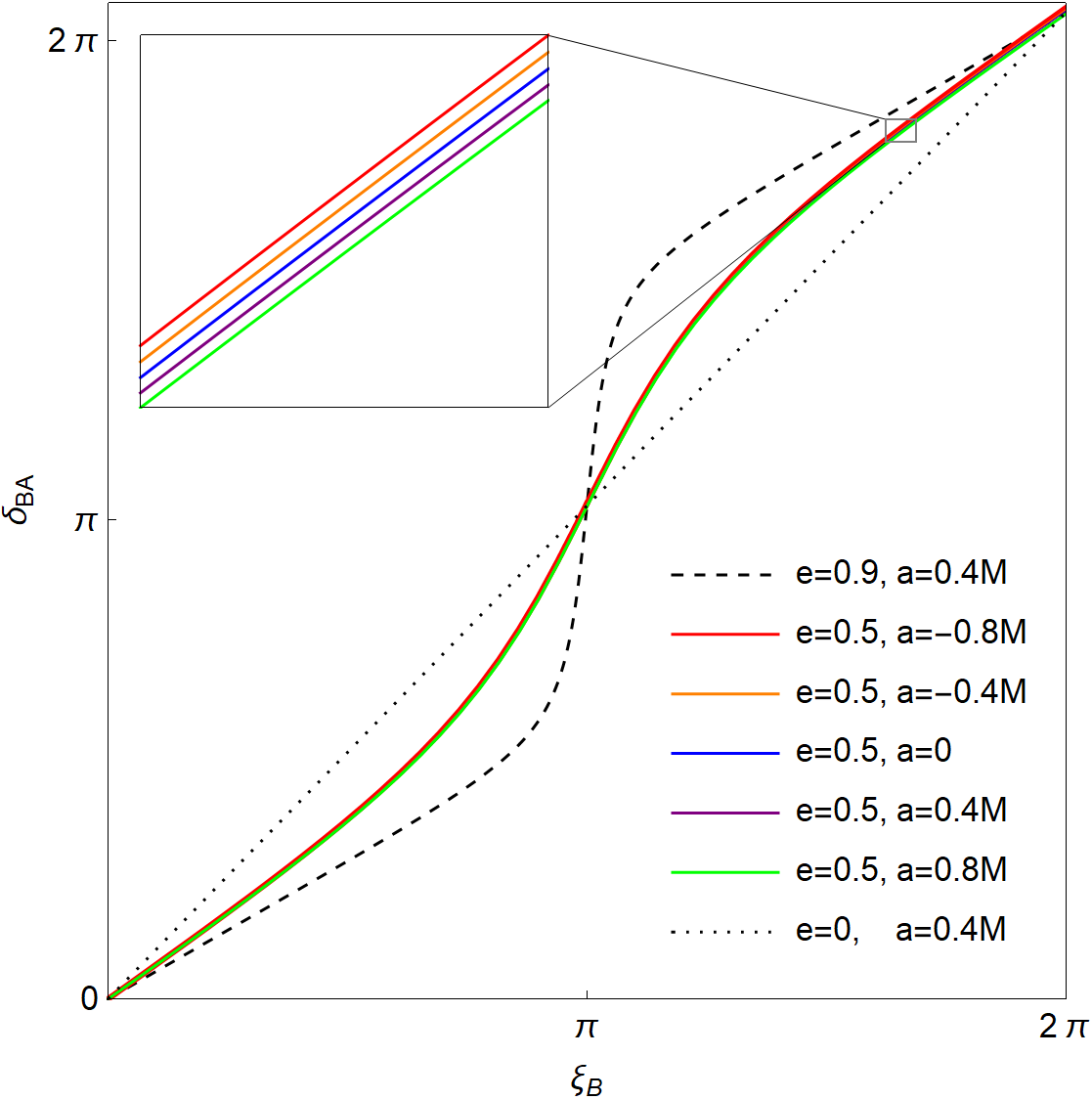} 
      \caption{$\delta_{BA}$ against $\chi_B$ with $\chi_A=0$, $p=10^2M$.}
  \label{fig-6}
\end{figure}

To visualize the parametric dependence of our deflection angle, we plot its behavior for selected parameter values. Fig.~\ref{fig-5} 
displays $\delta_{BA}$, $\phi_{BA}$, and $\Psi_{BA}$ as functions of $\chi_{B}$ (the decomposition $\delta_{BA}=\Psi_{BA}+\phi_{BA}$ is discussed in Sec.~\ref{GGWSAS}). Without loss of generality, we set $\xi_A=0$, namely, point $A$ is assumed to lie at a pericenter. The behavior of $\delta_{BA}$ is dominated by $\phi_{BA}$, while $\Psi_{BA}$ contributes a periodic modulation. $\phi_{BA}$ increases monotonically as point $B$ moves along the orbit, it slightly exceeds $2\pi$ when $B$ reaches the adjacent pericenter ($\chi_B=2\pi$), corresponding to the pericenter advance angle. $\Psi_{BA}$ varies periodically with a period of $2\pi$ and becomes zero when point $B$ reaches a pericenter or an apocenter, since the angle between the velocity and the outgoing radial direction is $\pi/2$ at both locations. Moreover, $\Psi_{BA}$ is more sensitive near the apocenter than near the pericenter. Fig.~\ref{fig-6}
illustrates the influence of the eccentricity $e$ and the spin parameter $a$ on $\delta_{BA}$. A larger $e$ leads to a greater absolute value of $\delta_{BA}$, i.e., a larger amplitude of its oscillation. For prograde orbits, a faster spin leads to a smaller $\delta_{BA}$, whereas for retrograde orbits, a faster spin leads to a larger $\delta_{BA}$.

\section{Conclusion}
\label{sec-5}
In this paper, we extend the GW method to bound orbits of massive particles in stationary spacetimes. The original GW method was designed to evaluate the deflection angle for unbound orbits in the asymptotic region, where both the source and the observer are assumed to be located at infinity. The subsequent introduction of the finite-distance deflection angle makes it possible to investigate orbit bending when the source and observer are situated at finite distances, which is more relevant to realistic astrophysical scenarios. Moreover, the proposal of the generalized GW method for stationary spacetimes permits arbitrary choice of the auxiliary circular arc, thereby substantially increasing the method’s flexibility and avoiding the restriction that bound orbits impose on the JMRF metric. Taking advantage of these recent developments, we establish a mapping between the bound orbit and the corresponding unbound orbit for massive particles, thereby enabling the GW method to be applied to a broader class of motions. To explicitly demonstrate the procedure and illustrate the feasibility of our approach, we compute the deflection angle between two arbitrary points along the bound orbit of a massive particle in Kerr spacetime and provide a brief analysis of the result.

\section*{Acknowledgments}
This work was supported in part by Natural Science Foundation of Hunan Province for Youths Grant No. 2024JJ6210 and 2024JJ6211, the National Natural Science Foundation of China Grant No. 12405054, 12375045, 12205093, and 12405053, and the Science Research Fund of Hunan Provincial Education Department No. 24B0476, 21A0297 and 24C0229.

\bibliographystyle{apsrev}
\bibliography{refs}

\end{document}